\documentclass[prd,showpacs,amsmath,amssymb,superscriptaddress,floatfix,nofootinbib,10pt]{revtex4}
\usepackage[utf8]{inputenc}
\usepackage[sort&compress]{natbib}
\usepackage{ulem}
\usepackage{bm}
\usepackage{times}
\usepackage{amssymb,amsbsy,amsmath,amsfonts}
\usepackage{graphicx}
\usepackage{float}
\usepackage{color}
\usepackage{morefloats}
\usepackage{rotating}
\usepackage{srcltx}
\usepackage{slashed}
\usepackage{subfigure}
\usepackage{multirow}
\usepackage{verbatim}
\usepackage{hyperref}
\usepackage{tabularx}

\begin{document}

\title{Strangeness $S = -3$ and $-4$ baryon-baryon interactions in relativistic chiral effective field theory}

\author{Zhi-Wei Liu}
\affiliation{School of Physics, Beihang University, Beijing, 102206, China}

\author{Jing Song}
\affiliation{School of Physics, Beihang University, Beijing, 102206, China}

\author{Kai-Wen Li}
\email[E-mail: ]{kaiwen.li@buaa.edu.cn}
\affiliation{Beijing Advanced Innovation Center for Big Data-Based Precision Medicine, School of Medicine and Engineering, Beihang University, Key Laboratory of Big Data-Based Precision Medicine (Beihang University), Ministry of Industry and Information Technology, Beijing, 100191, China}
\affiliation{School of Physics, Beihang University, Beijing, 102206, China}

\author{Li-Sheng Geng}
\email[E-mail: ]{lisheng.geng@buaa.edu.cn}
\affiliation{School of Physics, Beihang University, Beijing, 102206, China}
\affiliation{Beijing Key Laboratory of Advanced Nuclear Materials and Physics, Beihang University, Beijing, 102206, China}
\affiliation{Beijing Advanced Innovation Center for Big Data-Based Precision Medicine, School of Medicine and Engineering, Beihang University, Key Laboratory of Big Data-Based Precision Medicine (Beihang University), Ministry of Industry and Information Technology, Beijing, 100191, China}
\affiliation{School of Physics and Microelectronics, Zhengzhou University, Zhengzhou, Henan, 450001, China}

\begin{abstract}
The strangeness $S = -3$ and $-4$ baryon–baryon interactions are investigated in the relativistic chiral effective field theory at leading order. First, the 12 tree-level low-energy constants contributing to the $S = -1$ hyperon-nucleon interaction are fixed by fitting to the 36 hyperon-nucleon scattering data. Then the $S = -3$ and $-4$ baryon-baryon interactions are derived from that of $S = -1$ assuming that the corresponding low-energy constants are related to each other via SU(3) flavor symmetry. The comparison with the state-of-the-art lattice QCD simulations, show, however, that SU(3) flavor symmetry breaking effects can not be neglected. In order to take into account these effects, we redetermine two sets of low-energy constants by fitting to the lattice QCD data in the $\Xi\Sigma$ and $\Xi\Xi$ channels respectively. The fitting results demonstrate that the lattice QCD $S$-waves phase shifts for both channels can be described rather well. Without any additional free low-energy constants, the predicted phase shifts for the $^3D_1$ channel and the mixing angle $\varepsilon_1$ are also in qualitative agreement with the lattice QCD data for the $S = -3$ channel, while the results for the $S = -4$ channel remain to be checked by future lattice QCD simulations. With the so-obtained low-energy constants, the $S$-wave scattering lengths and effective ranges are calculated for these two channels at the physical point. Finally, in combination with the $S = 0$ and $-2$ results obtained in our previous works, we study the evolution of the irreducible representation ``$27$'' in the baryon-baryon interactions as a function of increasing strangeness. It is shown that the attraction decreases dramatically as strangeness increases from $S = 0$ to $S = -2$, but then remains relatively stable until $S = -4$. The results indicate that the existence of bound states in the $\Xi\Sigma$ and $\Xi\Xi$ channels is rather unlikely.

\end{abstract}

\pacs{13.75.Ev, 21.30.Fe, 12.39.Fe}

\maketitle

\section{Introduction}

Hyperon-nucleon ($YN$) and hyperon-hyperon ($YY$) interactions, as natural extensions of the nucleon-nucleon interaction in the $u$, $d$, $s$ flavor space, are fundamental quantities not only in hypernuclear physics but also in nuclear astrophysics~\cite{Nogga2002PhysRevLett88.172501, Lonardoni2015PhysRevLett114.092301}. There is no doubt that baryon–baryon interactions involving strangeness are facing an unprecedented opportunity with the development of large facilities for heavy-ion collisions and in the new era of multi-messenger astronomy~\cite{Jurgen2000PRL84.4305, Abbott2017PhysRevLett119.161101}.

Up to now, there is a fair amount of experimental data for the $\Lambda N$ and $\Sigma N$ systems~\cite{Sechi1968PhysRev75.1735, Alexander1968PhysRev173.1452, Eisele1971PhysLettB37.204, Engelmann1966PhysLett21.587, Hepp1968ZPhys214.71}, which have been used to determine the strangeness $S=-1$ hyperon-nucleon interaction~\cite{Polinder2006NPA779.244, KaiWen2016PhysRevD94.014029, KaiWen2018ChinPhysC42.014105}. However, direct data  are less stringent for the strangeness $S = -2$ sector~\cite{Takahashi2001PhysRevLett87.212502, Ahn2006PhysLettB633.214}, and even more so for the $S = -3$ and $-4$ systems.

In principle, SU(3) flavor symmetry could serve as a bridge to relate the strangeness $S = 0$ and $-1$ systems to those of $S = -3$ and $-4$~\cite{Stoks1999PhysRevC59.3009, Fujiwara2007ProgPartNuclPhys58.439, Haidenbauer2010PhysLettB684.275}. For instance, the Nijmegen group obtained the $S = -3$ and $-4$ $YY$ interactions in the one-boson-exchange model without any additional free parameters~\cite{Stoks1999PhysRevC59.3009}, based on (broken) SU(3) symmetry. The Kyoto-Niigata group investigated all the possible interactions between two octet baryons in the SU(6) quark cluster model, where SU(3) breaking was introduced via the Fermi–Breit interaction~\cite{Fujiwara2007ProgPartNuclPhys58.439}. Moreover, assuming strict SU(3) symmetry in the contact terms, the Bonn-J\"ulich group predicted the $S = -3$ and $-4$ $YY$ interactions from that of $S = -1$ in the non-relativistic chiral effective field theory (ChEFT) at leading order~\cite{Haidenbauer2010PhysLettB684.275}. It should be noted that due to lack of experimental constraints, substantial differences exist in the theoretical predictions for the $S = -3$ and $-4$ sectors, such as the existence of a $\Xi\Xi$ bound state and the magnitude of the $\Xi^0\Lambda$ cross section~\cite{Stoks1999PhysRevC59.3009, Fujiwara2007ProgPartNuclPhys58.439, Haidenbauer2010PhysLettB684.275}.

In this work, we use the relativistic ChEFT to study the $S = -3$ and $-4$ baryon-baryon interactions at leading order. This is an extension of our previous works on the $S=0,-1,-2$ baryon-baryon systems~\cite{Ren2018ChinPhysC42.014103, KaiWen2018ChinPhysC42.014105, Jing2018PhysRevC97.065201, KaiWen2018PhysRevC98.065203}. The leading order potential consists of four baryon contact terms (CT) and one-pseudoscalar-meson exchange (OPME) terms. $12$ and $5$ free low energy constants (LECs) appear in the CT potentials of the $S = -3$ and $-4$ sectors, respectively. First, we show that SU(3) flavor symmetry breaking must be taken into account in studying the $S = -3$ and $-4$ baryon-baryon interactions, in agreement with previous studies~\cite{Haidenbauer2015EPJA51.17,KaiWen2018ChinPhysC42.014105}. Next we use the latest lattice QCD (LQCD) results from the HAL QCD Collaboration~\cite{Ishii2018EPJWebConf175.05013, Doi2018EPJWebConf175.05009} to fix four of the $S$-wave LECs, independently for the $S = -3$ and $-4$ systems. The others would be left for future works. In addition, we also extrapolate the results to the physical point and study the systematics of SU(3) breaking from $NN$ to $\Xi\Xi$ interactions, particularly, the evolution of the ``$27$'' irreducible representation with increasing strangeness.

The paper is organized as follows. In Sec.~II we briefly introduce the relativistic ChEFT for the $S = -3$ and $-4$ baryon-baryon interactions. Results for the $\Xi\Sigma$ and $\Xi\Xi$ systems are shown in Sec. III. Finally, we conclude with a short summary and outlook in Sec. IV.

\section{Theoretical Framework}

In this section, the essential ingredients of the relativistic ChEFT will be briefly recalled for baryon-baryon interactions, especially for the strangeness $S = -3$ and $-4$ sectors at leading order (LO). For more details of the relativistic ChEFT, we refer the reader to Refs.~\cite{Ren2018ChinPhysC42.014103, KaiWen2018ChinPhysC42.014105, KaiWen2018PhysRevC98.065203, Jing2018PhysRevC97.065201}. In order to maintain Lorentz invariance, the complete baryon spinor is retained in the relativistic ChEFT approach, namely,
\begin{align}\label{Eq:Dirac spinor}
  u_B(\boldsymbol{p},s)=N_p\binom{1}{\frac{\boldsymbol{\sigma}\cdot\boldsymbol{p}}{E_p+M_B}}~\chi_s,\qquad N_p=\sqrt{\frac{E_p+M_B}{2M_B}},
\end{align}
where $E_p=\sqrt{\boldsymbol{p}^2+M_B^2}$, and $M_B$ is the averaged baryon mass, while a non-relativistic reduction of $u_B$ is employed in the non-relativistic ChEFT approach.

The LO baryon-baryon potentials consist of four-baryon contact terms (CT) without derivatives and one-pseudoscalar-meson exchange terms. The CT potential in momentum space reads,
\begin{align}\label{Eq:CPT}
  V_{\rm CT}^{\rm YY'}=C_i^{\rm YY'}\left(\bar{u}_3\Gamma_iu_1\right)\left(\bar{u}_4\Gamma_iu_2\right),
\end{align}
where $C_i^{\rm YY'}$ are the LECs corresponding to independent four-baryon terms. The superscript ${\rm YY'}$ denotes the hyperons in the reaction of $YN\rightarrow Y'N$. $\Gamma_i$ denote the elements of the Clifford algebra,
\begin{align}\label{Eq:Clifford_algebra}
  \Gamma_1=1,\qquad\Gamma_2=\gamma^\mu,\qquad\Gamma_3=\sigma^{\mu\nu},\qquad\Gamma_4=\gamma^\mu\gamma_5,\qquad\Gamma_5=\gamma_5.
\end{align}
The contact potentials are first calculated in the helicity basis, and then projected into different partial waves in the $|LSJ\rangle$ basis~\cite{Holzenkamp1989NPA500.485}. The partial wave projected potentials have the following form,
\begin{eqsize} 
  \begin{subequations}
    \begin{align}\label{Eq:CTP_LSJ}
      V_{\rm CT}^{\rm YY'}(^1S_0)
      &=\xi_B\left[\left(C_1^{\rm YY'}+C_2^{\rm YY'}-6C_3^{\rm YY'}+3C_4^{\rm YY'}\right)\left(1+R_p^2R_{p'}^2\right)+\left(3C_2^{\rm YY'}-6C_3^{\rm YY'}+C_4^{\rm YY'}+C_5^{\rm YY'}\right)\left(R_p^2+R_{p'}^2\right)\right]\nonumber\\
      &\equiv\xi_B\left[C_{1S0}^{\rm YY'}\left(1+R_p^2R_{p'}^2\right)+\hat{C}_{1S0}^{\rm YY'}\left(R_p^2+R_{p'}^2\right)\right],\\
      V_{\rm CT}^{\rm YY'}(^3S_1)
      &=\xi_B\left[\frac{1}{9}\left(C_1^{\rm YY'}+C_2^{\rm YY'}+2C_3^{\rm YY'}-C_4^{\rm YY'}\right)\left(9+R_p^2R_{p'}^2\right)+\frac{1}{3}\left(C_2^{\rm YY'}-2C_3^{\rm YY'}-C_4^{\rm YY'}-C_5^{\rm YY'}\right)\left(R_p^2+R_{p'}^2\right)\right]\nonumber\\
      &\equiv\xi_B\left[\frac{1}{9}C_{3S1}^{\rm YY'}\left(9+R_p^2R_{p'}^2\right)+\frac{1}{3}\hat{C}_{3S1}^{\rm YY'}\left(R_p^2+R_{p'}^2\right)\right],\\
      V_{\rm CT}^{\rm YY'}(^3P_0)
      &=\xi_B\left[-2\left(C_1^{\rm YY'}-4C_2^{\rm YY'}-12C_3^{\rm YY'}-4C_4^{\rm YY'}+C_5^{\rm YY'}\right)R_pR_{p'}\right]\equiv\xi_B\left[-2C_{3P0}^{\rm YY'}R_pR_{p'}\right],\\
      V_{\rm CT}^{\rm YY'}(^3P_1)
      &=\xi_B\left[-\frac{4}{3}\left(C_1^{\rm YY'}-2C_2^{\rm YY'}+2C_4^{\rm YY'}-C_5^{\rm YY'}\right)R_pR_{p'}\right]=\xi_B\left[-\frac{4}{3}\left(C_{1S0}^{\rm YY'}-\hat{C}_{1S0}^{\rm YY'}\right)R_pR_{p'}\right],\\
      V_{\rm CT}^{\rm YY'}(^1P_1)
      &=\xi_B\left[-\frac{2}{3}\left(C_1^{\rm YY'}+4C_3^{\rm YY'}+C_5^{\rm YY'}\right)R_pR_{p'}\right]=\xi_B\left[-\frac{2}{3}\left(C_{3S1}^{\rm YY'}-\hat{C}_{3S1}^{\rm YY'}\right)R_pR_{p'}\right],\\
      V_{\rm CT}^{\rm YY'}(^3D_1)
      &=\xi_B\left[\frac{8}{9}\left(C_1^{\rm YY'}+C_2^{\rm YY'}+2C_3^{\rm YY'}-C_4^{\rm YY'}\right)R_p^2R_{p'}^2\right]=\xi_B\left[\frac{8}{9}C_{3S1}^{\rm YY'}R_p^2R_{p'}^2\right],\\
      V_{\rm CT}^{\rm YY'}(^3SD_1)
      &=\xi_B\left[\frac{2\sqrt{2}}{9}\left(C_1^{\rm YY'}+C_2^{\rm YY'}+2C_3^{\rm YY'}-C_4^{\rm YY'}\right)R_p^2R_{p'}^2+\frac{2\sqrt{2}}{3}\left(C_2^{\rm YY'}-2C_3^{\rm YY'}-C_4^{\rm YY'}-C_5^{\rm YY'}\right)R_p^2\right]\nonumber\\
      &=\xi_B\left[\frac{2\sqrt{2}}{9}C_{3S1}^{\rm YY'}R_p^2R_{p'}^2+\frac{2\sqrt{2}}{3}\hat{C}_{3S1}^{\rm YY'}R_p^2\right],\\
      V_{\rm CT}^{\rm YY'}(^3DS_1)
      &=\xi_B\left[\frac{2\sqrt{2}}{9}\left(C_1^{\rm YY'}+C_2^{\rm YY'}+2C_3^{\rm YY'}-C_4^{\rm YY'}\right)R_p^2R_{p'}^2+\frac{2\sqrt{2}}{3}\left(C_2^{\rm YY'}-2C_3^{\rm YY'}-C_4^{\rm YY'}-C_5^{\rm YY'}\right)R_{p'}^2\right]\nonumber\\
      &=\xi_B\left[\frac{2\sqrt{2}}{9}C_{3S1}^{\rm YY'}R_p^2R_{p'}^2+\frac{2\sqrt{2}}{3}\hat{C}_{3S1}^{\rm YY'}R_{p'}^2\right],
    \end{align}
  \end{subequations}
\end{eqsize}where $\xi_B = N_p^2N_{p'}^2,~R_p = |\boldsymbol{p}|/(E_p+M_B),~R_{p'} = |\boldsymbol{p'}|/(E_{p'}+M_B)$. $\boldsymbol{p}$ and $\boldsymbol{p'}$ denote the initial and final momenta, respectively. Here, we only list the final results for independent potentials respecting SU(3) symmetry, as shown in Table~\ref{Tab:SU3}. There are 12 independent LECs in the present work, namely, $C_{1S0}^{\Lambda\Lambda}$, $C_{1S0}^{\Sigma\Sigma}$, $C_{3S1}^{\Lambda\Lambda}$, $C_{3S1}^{\Sigma\Sigma}$, $C_{3S1}^{\Lambda\Sigma}$, $\hat{C}_{1S0}^{\Lambda\Lambda}$, $\hat{C}_{1S0}^{\Sigma\Sigma}$, $\hat{C}_{3S1}^{\Lambda\Lambda}$, $\hat{C}_{3S1}^{\Sigma\Sigma}$, $\hat{C}_{3S1}^{\Lambda\Sigma}$, $C_{3P0}^{\Lambda\Lambda}$ and $C_{3P0}^{\Sigma\Sigma}$. 
\begin{table}[ht]
  \centering
  \caption{Leading order baryon–baryon contact potentials in the isospin basis.}
  \label{Tab:SU3}
  \setlength{\tabcolsep}{23.8pt}
  \begin{tabular}{lccll}
  \hline
	\hline
	\multicolumn{1}{c}{\multirow{2}{*}{}}&\multicolumn{1}{c}{\multirow{2}{*}{Channel}}&
	\multicolumn{1}{c}{\multirow{2}{*}{Isospin}}&\multicolumn{2}{c}{$V_{\rm CT}$}\\
	\cline{4-5}
	&&&$\xi=~^1S_0,^3P_0,^3P_1$ &$\zeta=~^3S_1,^1P_1,^3D_1,^3SD_1$\\
	\hline
	$S=0$  &$NN\rightarrow NN$ &0 &/  &$V_{\zeta}^{\Lambda\Lambda}+V_{\zeta}^{\Lambda\Sigma}$\\
		     &$NN\rightarrow NN$ &1 &$V_{\xi}^{\Sigma\Sigma}$ &/ \\
	\hline
	$S=-1$ &$\Lambda N\rightarrow\Lambda N$ &$1/2$ &$V_{\xi}^{\Lambda\Lambda}$ &$V_{\zeta}^{\Lambda\Lambda}$\\
		     &$\Lambda N\rightarrow\Sigma N$ &$1/2$ &$3(V_{\xi}^{\Lambda\Lambda}-V_{\xi}^{\Sigma\Sigma})$ &$V_{\zeta}^{\Lambda\Sigma}$\\
		     &$\Sigma N\rightarrow\Sigma N$ &$1/2$ &$9V_{\xi}^{\Lambda\Lambda}-8V_{\xi}^{\Sigma\Sigma}$ &$V_{\zeta}^{\Lambda\Lambda}$\\
		     &$\Sigma N\rightarrow\Sigma N$ &$3/2$ &$V_{\xi}^{\Sigma\Sigma}$ &$V_{\zeta}^{\Sigma\Sigma}$\\
  \hline
  $S=-2$ &$\Sigma\Sigma\rightarrow\Sigma\Sigma$ &2 &$V_{\xi}^{\Sigma\Sigma}$ &/\\
         &\dots\\
  \hline
	$S=-3$ &$\Xi\Lambda\rightarrow\Xi\Lambda$ &$1/2$ &$V_{\xi}^{\Lambda\Lambda}$ &$(V_{\zeta}^{\Sigma\Sigma}+V_{\zeta}^{\Lambda\Lambda}-V_{\zeta}^{\Lambda\Sigma})/2$\\
		     &$\Xi\Lambda\rightarrow\Xi\Sigma$ &$1/2$ &$3(V_{\xi}^{\Lambda\Lambda}-V_{\xi}^{\Sigma\Sigma})$ &$(V_{\zeta}^{\Sigma\Sigma}-V_{\zeta}^{\Lambda\Lambda}+V_{\zeta}^{\Lambda\Sigma})/2$\\
		     &$\Xi\Sigma\rightarrow\Xi\Sigma$ &$1/2$ &$9V_{\xi}^{\Lambda\Lambda}-8V_{\xi}^{\Sigma\Sigma}$ &$(V_{\zeta}^{\Sigma\Sigma}+V_{\zeta}^{\Lambda\Lambda}-V_{\zeta}^{\Lambda\Sigma})/2$\\
		     &$\Xi\Sigma\rightarrow\Xi\Sigma$ &$3/2$ &$V_{\xi}^{\Sigma\Sigma}$ &$V_{\zeta}^{\Lambda\Lambda}+V_{\zeta}^{\Lambda\Sigma}$\\
	\hline
	$S=-4$ &$\Xi\Xi\rightarrow\Xi\Xi$ &0 &/ &$V_{\zeta}^{\Sigma\Sigma}$\\
		     &$\Xi\Xi\rightarrow\Xi\Xi$ &1 &$V_{\xi}^{\Sigma\Sigma}$ &/\\
	\hline
	\hline
  \end{tabular}
\end{table}

The OPME potentials in momentum space can be written as,
\begin{align}\label{Eq:OPMEP}
  V_{\rm OPME}=-N_{B_1B_3\phi}N_{B_2B_4\phi}\frac{\left(\bar{u}_3\gamma^\mu\gamma_5q_\mu u_1\right)\left(\bar{u}_4\gamma^\nu\gamma_5q_\nu u_2\right)}{q^2-m^2}\mathcal{I}_{B_1B_2\rightarrow B_3B_4},
\end{align}
where $q = p' - p$ is the momentum transfer, $q^2 = (E_{p'} - E_p)^2 - (\boldsymbol{p}' - \boldsymbol{p})^2$, and $m$ is the mass of the exchanged pseudoscalar meson. The SU(3) coefficient $N_{BB'\phi}$ and isospin factor $\mathcal{I}_{B_1B_2\rightarrow B_3B_4}$ can be found in Refs.~\cite{Polinder2006NPA779.244, KaiWen2016PhysRevD94.014029}. It is easy to obtain $V_{\rm OPME}$ in the $|LSJ\rangle$ basis following the same procedure as that for the contact terms. Note that due to the mass difference of exchanged mesons, SU(3) symmetry is not fulfilled strictly in the OPME potentials. 

To take into account the non-perturbative nature of the baryon-baryon interactions, following Ref.~\cite{KaiWen2016PhysRevD94.014029}, we solve the coupled-channel Kadyshevsky equation,
\begin{align}\label{Eq:Kadyshevsky}
  T_{\rho\rho'}^{\nu\nu',J}(\boldsymbol{p}',\boldsymbol{p};\sqrt{s})=V_{\rho\rho'}^{\nu\nu',J}(\boldsymbol{p}',\boldsymbol{p})+\sum_{\rho^{\prime\prime},\nu^{\prime\prime}}\int_0^\infty\frac{{\rm d}p^{\prime\prime}p^{\prime\prime2}}{(2\pi)^3}\frac{M_{B_{1,\nu^{\prime\prime}}}M_{B_{2,\nu^{\prime\prime}}}V_{\rho\rho^{\prime\prime}}^{\nu\nu^{\prime\prime},J}(\boldsymbol{p}',\boldsymbol{p}^{\prime\prime})T_{\rho^{\prime\prime}\rho'}^{\nu^{\prime\prime}\nu',J}(\boldsymbol{p}^{\prime\prime},\boldsymbol{p};\sqrt{s})}{E_{1,\nu^{\prime\prime}}E_{2,\nu^{\prime\prime}}(\sqrt{s}-E_{1,\nu^{\prime\prime}}-E_{2,\nu^{\prime\prime}}+i\epsilon)},
\end{align}
where $\sqrt{s}$ is the total energy of the baryon-baryon system in the center-of-mass frame and $E_{n,\nu''} = \sqrt{\boldsymbol{p}''^2 + M_{B_{n,\nu''}}^2}$, ($n = 1, 2$). The labels $\nu$, $\nu'$, $\nu''$ denote the particle channels, and $\rho$, $\rho'$, $\rho''$ denote the partial waves. In addition, to avoid ultraviolet divergence in numerical evaluations, baryon-baryon potentials are regularized with an exponential form factor,
\begin{align}\label{Eq:regularization}
  f_{\Lambda_F}(\boldsymbol{p},\boldsymbol{p'})=\exp\left[-\left(\frac{\boldsymbol{p}}{\Lambda_F}\right)^{2n}-\left(\frac{\boldsymbol{p'}}{\Lambda_F}\right)^{2n}\right],
\end{align}
where $n = 2$~\cite{Epelbaum2005NPA747.362}. In the present work, following Ref.~\cite{Haidenbauer2010PhysLettB684.275}, we consider cutoff values in the range of $550 – 700$ MeV.

\section{Results and Discussion}

\subsection{Predictions for the $S = -3$ and $-4$ baryon–baryon interactions via strict SU(3) symmetry}

As described in the previous section, the $12$ LECs appearing in the $S = -3$ and $-4$ systems are the same as those in the $S = -1$ sector~\cite{KaiWen2018ChinPhysC42.014105}, assuming strict SU(3) symmetry. In order to be self-consistent, we have refitted the LECs to the 36 $S = -1$ $YN$ scattering data~\cite{Sechi1968PhysRev75.1735, Alexander1968PhysRev173.1452, Eisele1971PhysLettB37.204, Engelmann1966PhysLett21.587,Hepp1968ZPhys214.71} with the average baryon mass $M_B = 1151$ MeV, instead of $M_B = 1080$ MeV as in our previous work~\cite{KaiWen2018ChinPhysC42.014105}\footnote{$M_B = 1080$ MeV is obtained by taking the average mass of $N$, $\Lambda$, and $\Sigma$ baryons, while $M_B = 1151$ MeV is the average mass of the octet baryons.}. In addition, we choose $C_{3P0}^{\Lambda\Lambda}$ and $C_{3P0}^{\Sigma\Sigma}$ to be the $P$-wave free parameters in the fits. The updated values of the LECs for different $\Lambda_F$ are listed in Table~\ref{Tab:LECs_m1}.

\begin{table}[h]
  \centering
  \caption{Low-energy constants (in units of $10^4~{\rm GeV}^{-2}$) obtained with  various cutoff $\Lambda_F$ (in units of MeV) in the relativistic ChEFT. These LECs are determined by fitting to the $S = -1$ hyperon-nucleon scattering data.}
  \label{Tab:LECs_m1}
  \setlength{\tabcolsep}{5.2pt}
  \begin{tabular}{cccccccccrcccc}
    \hline
    \hline
    $\Lambda_F$&&$C_{1S0}^{\Lambda\Lambda}$ &$C_{1S0}^{\Sigma\Sigma}$ &$C_{3S1}^{\Lambda\Lambda}$ &$C_{3S1}^{\Sigma\Sigma}$ &$C_{3S1}^{\Lambda\Sigma}$ &$\hat{C}_{1S0}^{\Lambda\Lambda}$ &$\hat{C}_{1S0}^{\Sigma\Sigma}$ &$\hat{C}_{3S1}^{\Lambda\Lambda}$~~ &$\hat{C}_{3S1}^{\Sigma\Sigma}$ &$\hat{C}_{3S1}^{\Lambda\Sigma}$ &$C_{3P0}^{\Lambda\Lambda}$ &$C_{3P0}^{\Sigma\Sigma}$\\
    \hline
    $550$  &&$-0.0671$ &$-0.0951$ &$0.0244$ &$0.0696$ &$0.0528$ &$3.0342$ &$3.3680$ &$ 1.0971$ &$-0.2827$ &$1.5582$ &$-2.7564$ &$-1.2394$ \\
    $600$  &&$-0.0553$ &$-0.0801$ &$0.0244$ &$0.0839$ &$0.0384$ &$3.0928$ &$3.4223$ &$ 0.5519$ &$-0.2351$ &$1.2292$ &$-2.7674$ &$-1.3346$ \\
    $650$  &&$-0.0377$ &$-0.0588$ &$0.0255$ &$0.0995$ &$0.0254$ &$3.1119$ &$3.4313$ &$ 0.1908$ &$-0.2344$ &$0.9751$ &$-2.7698$ &$-1.4623$ \\
    $700$  &&$-0.0126$ &$-0.0296$ &$0.0293$ &$0.1163$ &$0.0168$ &$3.1250$ &$3.4343$ &$-0.0179$ &$-0.2273$ &$0.7962$ &$-2.7703$ &$-1.6516$ \\
    \hline
    \hline
  \end{tabular}
\end{table}

More recently, the HAL QCD Collaboration have performed LQCD simulations for the $S = -3$ and $-4$ baryon-baryon interactions with an almost physical pion mass ($m_\pi = 146$ MeV), using the so-called HAL QCD approach. They have obtained time-dependent $S$- and $D$-wave phase shifts for $\Xi\Sigma$ ($I = 3/2$) scattering with time $t = 11 - 15$~\cite{Ishii2018EPJWebConf175.05013} and $S$-wave phase shifts for $\Xi\Xi$ scattering with $t = 16 - 18$~\cite{Doi2018EPJWebConf175.05009}. These data provide us an opportunity to check the above obtained ChEFT potential. As the LQCD results are still obtained at unphysical pion masses, though very close to the physical point, we employ the LQCD hadron masses in our numerical study, i.e., $m_\pi = 146$ MeV, $m_K = 525$ MeV, $M_\Xi = 1356$ MeV, $M_\Sigma = 1222$ MeV, and the average mass of the octet baryons $M_B = 1179$ MeV~\cite{Ishii2018EPJWebConf175.05013, Doi2018EPJWebConf175.05009}. 

\begin{figure}[h]
  \centering
  \includegraphics[width=0.7\textwidth]{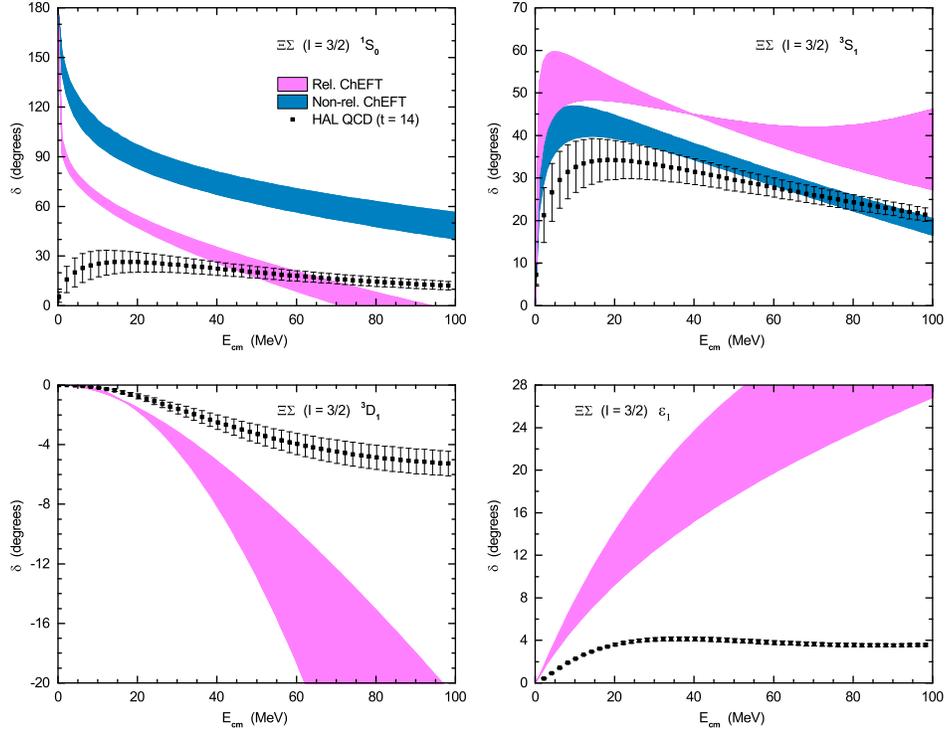}
  \caption{(color online) $\Xi\Sigma$ ($I = 3/2$) phase shifts for the $^1S_0$, $^3S_1$, $^3D_1$ channels and the mixing angle $\varepsilon_1$ as functions of the c.m. kinetic energy. The results are calculated in the relativistic ChEFT (light Magenta) and the non-relativistic ChEFT (dark blue) at LO. The shaded bands reflect the variation of the cutoff in the range $\Lambda_F = 550 – 700$ MeV. The HAL QCD phase shifts ($t = 14$) are from Ref.~\cite{Ishii2018EPJWebConf175.05013}.  Note that SU(3) symmetry has been employed in relating the relevant LECs to those of the $S = -1$ sector.
  }\label{Fig:PS_XiSigma_prediction}
\end{figure}

In Fig.~\ref{Fig:PS_XiSigma_prediction} we compare the $\Xi\Sigma$ ($I = 3/2$) phase shifts for the $^1S_0$, $^3S_1$, $^3D_1$ partial waves and the mixing angle $\varepsilon_1$ with the HAL QCD results ($t = 14$)~\cite{Ishii2018EPJWebConf175.05013}.
It is worthwhile to emphasize that the $^3S_1$, $^3D_1$ channels and the mixing angle $\varepsilon_1$ can be described self-consistently in the relativistic ChEFT at LO. From Fig.~\ref{Fig:PS_XiSigma_prediction} one observes that both the relativistic and non-relativistic ChEFT predict strong attractions in the $\Xi\Sigma$ ($I = 3/2$) $^1S_0$ channel, suggesting the likely existence of a bound state. On the other hand, though the interaction given by the HAL QCD Collaboration is attractive, the strength is not strong enough to generate a bound state. For the $^3S_1-{}^3D_1$ channels, there are still quantitative differences between the ChEFT results and the HAL QCD results, especially for the $^3D_1$ channel and the mixing angle $\varepsilon_1$. These discrepancies indicate that SU(3) flavor symmetry breaking must be taken into account in relating the $S = -1$ and $S = -3$ baryon-baryon interactions. 

\begin{figure}[h]
  \centering
  \includegraphics[width=0.7\textwidth]{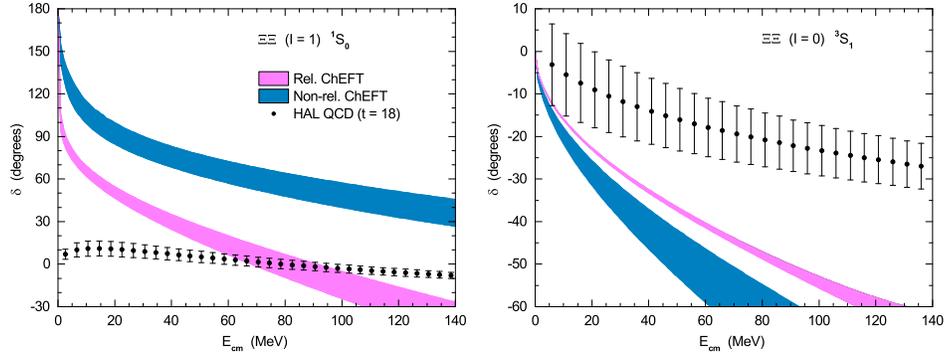}
  \caption{(color online) $\Xi\Xi$ phase shifts for the $^1S_0$ and $^3S_1$ channels as functions of the c.m. kinetic energy. The theoretical results are calculated in the relativistic ChEFT (light Magenta) and the non-relativistic ChEFT (dark blue) at LO. The shaded bands reflect the variation of the cutoff in the range $\Lambda_F = 550 – 700$ MeV. The HAL QCD phase shifts ($t = 18$) are from Ref.~\cite{Doi2018EPJWebConf175.05009}. Note that SU(3) symmetry has been employed in relating the relevant LECs to those of the $S = -1$ sector.}\label{Fig:PS_XiXi_prediction}
\end{figure}

For the $S = -4$ sector, the HAL QCD Collaboration have studied the two-$\Xi$ systems for the $^1S_0$ channel and $^3S_1-{}^3D_1$ coupled channel, while they only showed the phase shifts for the $^1S_0$ and $^3S_1$ partial waves in Ref.~\cite{Doi2018EPJWebConf175.05009}. Accordingly, we also calculate the $\Xi\Xi$ phase shifts for the $^1S_0$ and $^3S_1$ channels in the relativistic ChEFT and the non-relativistic ChEFT. As shown in Fig.~\ref{Fig:PS_XiXi_prediction}, there exist relatively large discrepancies between the ChEFT results and the HAL QCD results for the $\Xi\Xi$ phase shifts, indicating the need to take into account SU(3) breaking effects.

From the above comparisons with the LQCD results, we conclude that SU(3) symmetry breaking is large when one moves from the $S=-1$ system to the $S=-3,-4$ sectors. As a result, we refrain from using the LECs tabulated in Table~\ref{Tab:LECs_m1} to predict the corresponding cross sections for these two sectors.

\subsection{Fits to the $S = -3$ and $-4$ LQCD data}

As shown in the previous subsection, SU(3) flavor symmetry breaking effects must be taken into account in relating the $S = -3$ and $-4$ baryon-baryon interactions with those of $S = -1$. In practical terms, this implies that we have to refit the relevant LECs to the LQCD data.

First we fit to the $\Xi\Sigma$ ($I = 3/2$) $S$-wave phase shifts with the center-of-mass energy $E_{\textrm{cm}}\leq 30$ MeV~\cite{Ishii2018EPJWebConf175.05013}. As it has been described in Ref.~\cite{Doi2018EPJWebConf175.05009}, the present HAL QCD approach could provide more reliable results with increasing time $t$, but the uncertainties increase as well. To balance reliability and uncertainty, we studied the phase shifts obtained with $t = 12, 13, 14$ and found that the results of $t = 14$ can be better described in the whole energy region. Therefore, in the following, we only focus on the LQCD results obtained with $t = 14$. Six LECs appear in this single channel and the corresponding potentials read,
\begin{subequations}
  \begin{align}
    V_{\rm CT,~I=3/2}^{\rm\Xi\Sigma\rightarrow\Xi\Sigma}(^1S_0)
    &=\xi_B\left[C_{1S0}^{\rm\Sigma\Sigma}\left(1+R_p^2R_{p'}^2\right)+\hat{C}_{1S0}^{\rm\Sigma\Sigma}\left(R_p^2+R_{p'}^2\right)\right],\\
    V_{\rm CT,~I=3/2}^{\rm\Xi\Sigma\rightarrow\Xi\Sigma}(^3S_1)
    &=\xi_B\left[\frac{1}{9}\left(C_{3S1}^{\rm\Lambda\Lambda}+C_{3S1}^{\rm\Lambda\Sigma}\right)\left(9+R_p^2R_{p'}^2\right)+\frac{1}{3}\left(\hat{C}_{3S1}^{\rm\Lambda\Lambda}+\hat{C}_{3S1}^{\rm\Lambda\Sigma}\right)\left(R_p^2+R_{p'}^2\right)\right].
  \end{align}
\end{subequations}
Note that in the $^3S_1$ partial wave only the two combinations, namely, $C_{3S1}^{\Lambda\Lambda} + C_{3S1}^{\Lambda\Sigma}$ and $\hat{C}_{3S1}^{\Lambda\Lambda} + \hat{C}_{3S1}^{\Lambda\Sigma}$ can be determined. The cutoff value $\Lambda_F$ is varied in the range of $550 - 700$ MeV. Then we extrapolate the results to the physical point. It is noted that the phase shifts of $\Xi\Sigma$ $^3D_1$ and $\varepsilon_1$ from LQCD are not used in the fits, which will be discussed in the following.

A similar strategy is applied to the fits of the $\Xi\Xi$ system. The $S$-wave LQCD data with the same energy range but different time ($t = 18$) are taken into account. Four LECs in the CT potential need to be determined, which are defined as,
\begin{subequations}
  \begin{align}
    V_{\rm CT,~I=1}^{\rm\Xi\Xi\rightarrow\Xi\Xi}(^1S_0)
    &=\xi_B\left[C_{1S0}^{\rm\Sigma\Sigma}\left(1+R_p^2R_{p'}^2\right)+\hat{C}_{1S0}^{\rm\Sigma\Sigma}\left(R_p^2+R_{p'}^2\right)\right],\\
    V_{\rm CT,~I=0}^{\rm\Xi\Xi\rightarrow\Xi\Xi}(^3S_1)
    &=\xi_B\left[\frac{1}{9}C_{3S1}^{\rm\Sigma\Sigma}\left(9+R_p^2R_{p'}^2\right)+\frac{1}{3}\hat{C}_{3S1}^{\rm\Sigma\Sigma}\left(R_p^2+R_{p'}^2\right)\right],
  \end{align}
\end{subequations}
where $C_{1S0}^{\Sigma\Sigma}$ and $\hat{C}_{1S0}^{\Sigma\Sigma}$ are the same as those in the $S = -3$ system, as they belong to the same  SU(3) irreducible representation ``$27$''.

\begin{figure}[h]
  \centering
  \includegraphics[width=0.7\textwidth]{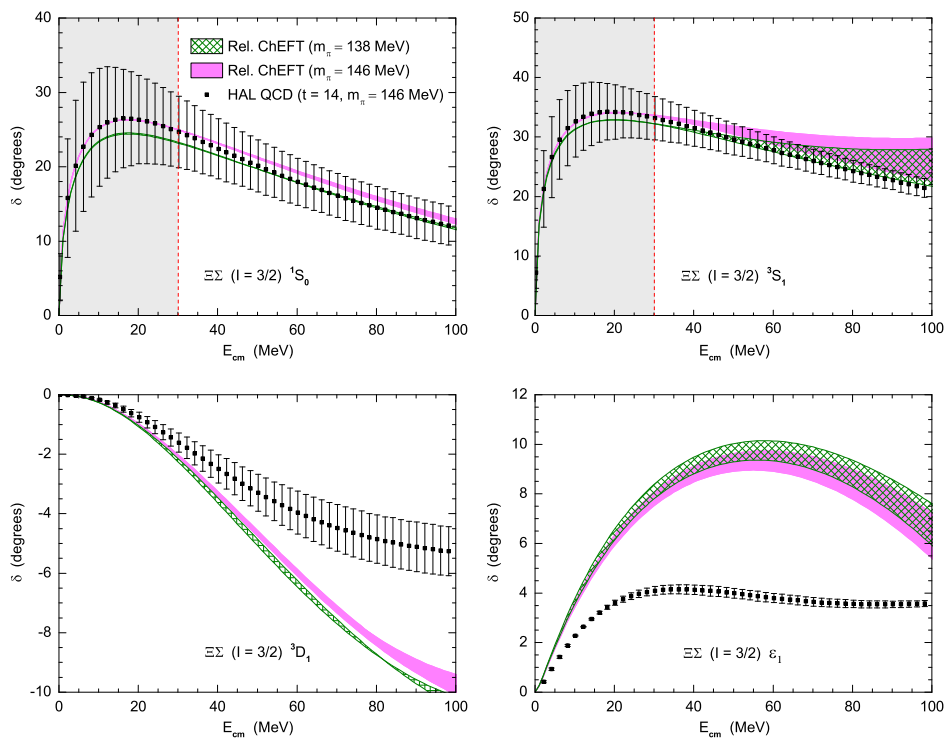}
  \caption{(color online) Phase shifts for the $\Xi\Sigma$ ($I = 3/2$) $S$- and $D$-wave as functions of the c.m. kinetic energy. The LECs are fitted to the $^1S_0$ and $^3S_1$ phase shifts from the HAL QCD Collaboration ($t = 14$)~\cite{Ishii2018EPJWebConf175.05013}, for energies up to 30 MeV. The shaded bands show the variation of the cutoff in the range $\Lambda$ = 550 – 700 MeV.
  }\label{Fig:PS_XiSigma_fitting}
\end{figure}

The fitted and extrapolated $\Xi\Sigma$ ($I = 3/2$) $S$- and $D$-wave phase shifts are shown in Fig.~\ref{Fig:PS_XiSigma_fitting}. The bands represent the variations within the cutoff range of $\Lambda_F = 550 - 700$ MeV. The corresponding values of the LECs are listed in Table~\ref{Tab:LECs_m3}. The two $S$-wave phase shifts are in very good agreement with the LQCD data, while the predicted $^3D_1$ phase shifts and $\varepsilon_1$  are also qualitatively similar to the LQCD data but are larger than their LQCD counterparts at high energies. The reason might be traced back to the fact that at LO, the same two LECs are responsible for the $^3S_1-{}^3D_1$ coupled channels. 

\begin{table}[h]
  \centering
  \caption{Low-energy constants (in units of $10^4~{\rm GeV}^{-2}$) for various cutoff $\Lambda_F$ (in units of MeV) in the relativistic ChEFT. These LECs are determined by fitting to the $\Xi\Sigma$ $^1S_0$ and $^3S_1$ phase shifts up to $E_{\rm cm} = 30$ MeV, taken from the HAL QCD Collaboration ($t = 14$)~\cite{Ishii2018EPJWebConf175.05013}.}
  \label{Tab:LECs_m3}
  \setlength{\tabcolsep}{27.5pt}
  \begin{tabular}{ccccccc}
    \hline
    \hline
    \multicolumn{1}{c}{\multirow{2}{*}{$\Lambda_F$}}&\multicolumn{5}{c}{LECs\quad($S = -3$)}\\
    \cline{3-6}
    &&$C_{1S0}^{\Sigma\Sigma}$ &$C_{3S1}^{\Lambda\Lambda}+C_{3S1}^{\Lambda\Sigma}$ &$\hat{C}_{1S0}^{\Sigma\Sigma}$ &$\hat{C}_{3S1}^{\Lambda\Lambda}+\hat{C}_{3S1}^{\Lambda\Sigma}$ \\
    \hline
    $550$  &&$-0.0349$ &$-0.0315$ &$-0.0875$ &$-0.8322$ \\
    $600$  &&$-0.0348$ &$-0.0294$ &$-0.0677$ &$-0.8514$ \\
    $650$  &&$-0.0347$ &$-0.0278$ &$-0.0555$ &$-0.8855$ \\
    $700$  &&$-0.0347$ &$-0.0267$ &$-0.0474$ &$-0.9126$ \\
    \hline
    \hline
  \end{tabular}
\end{table}

\begin{figure}[h]
  \centering
  \includegraphics[width=0.7\textwidth]{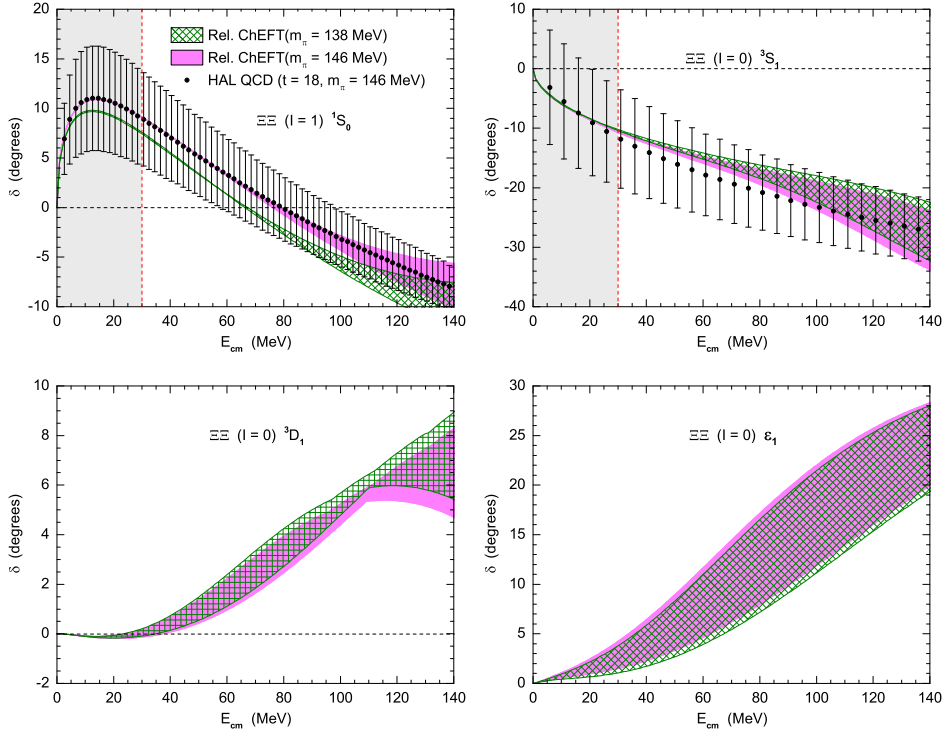}
  \caption{(color online) Phase shifts for the $\Xi\Xi$ ($I = 1$) $^1S_0$ and ($I = 0$) $^3S_1$ partial waves as functions of the c.m. kinetic energy. The LECs are fitted  to the $^1S_0$ and $^3S_1$ phase shifts provided by the HAL QCD Collaboration ($t = 18$)~\cite{Doi2018EPJWebConf175.05009}, for energies up to $30$  MeV. The shaded bands show the variation of the cutoff in the range $\Lambda_F = 550 - 700$ MeV. 
  }\label{Fig:PS_XiXi_fitting}
\end{figure}

For the $\Xi\Xi$ system, we show the fitted and extrapolated results in Fig.~\ref{Fig:PS_XiXi_fitting} within the same cutoff range. The corresponding LECs are listed in Table~\ref{Tab:LECs_m4}. The differences of the LECs values listed in Tables~\ref{Tab:LECs_m1}, \ref{Tab:LECs_m3}, and \ref{Tab:LECs_m4}, especially for the $\hat{C}_{1S0}^{\Sigma\Sigma}$, testify the SU(3) flavor symmetry breaking in different strangeness sectors again. The relativistic ChEFT can describe the $S$-wave LQCD data very well. Phase shifts of the $^3D_1$ and $\varepsilon_1$ are also predicted, though no LQCD data exist. The mixing angle $\varepsilon_1$ of the $\Xi\Xi$ system is even larger than that of the $\Xi\Sigma$ system, which implies that the $^3S_1-{}^3D_1$ coupling is even stronger. Given the comparison in the $S = -3$ system, we anticipate that the real numbers might be smaller but the trend should be the same, namely, positive phase shifts for $^3D_1$ and positive $\varepsilon_1$, which can be tested by future LQCD simulations.

\begin{table}[h]
  \centering
  \caption{Low-energy constants (in units of $10^4~{\rm GeV}^{-2}$) at various cutoff $\Lambda_F$ (in units of MeV) in the relativistic ChEFT. These LECs are determined by fitting to the $\Xi\Xi$ $^1S_0$ and $^3S_1$ phase shifts up to $E_{\rm cm} = 30$ MeV, taken from the HAL QCD Collaboration ($t = 18$)~\cite{Doi2018EPJWebConf175.05009}.}
  \label{Tab:LECs_m4}
  \setlength{\tabcolsep}{31.5pt}
  \begin{tabular}{cccccc}
    \hline
    \hline
    \multicolumn{1}{c}{\multirow{2}{*}{$\Lambda_F$}}&\multicolumn{5}{c}{LECs\quad($S = -4$)}\\
    \cline{3-6}
    &&$C_{1S0}^{\Sigma\Sigma}$ &$C_{3S1}^{\Sigma\Sigma}$ &$\hat{C}_{1S0}^{\Sigma\Sigma}$ &$\hat{C}_{3S1}^{\Sigma\Sigma}$\\
    \hline
    $550$  &&$-0.0221$ &$0.0195$ &$-0.0356$ &$1.3522$ \\
    $600$  &&$-0.0221$ &$0.0193$ &$-0.0267$ &$1.1246$ \\
    $650$  &&$-0.0220$ &$0.0193$ &$-0.0197$ &$0.9594$ \\
    $700$  &&$-0.0218$ &$0.0191$ &$-0.0140$ &$0.8279$ \\
    \hline
    \hline
  \end{tabular}
\end{table}

The extrapolated phase shifts to the physical point for both the $S = -3$ and $-4$ baryon-baryon interactions, shown in Figs.~\ref{Fig:PS_XiSigma_fitting} and \ref{Fig:PS_XiXi_fitting}, are almost the same as the fitted results. This is reasonable since the LQCD simulations were performed with almost physical hadron masses. Moreover, we also calculate the physical scattering lengths and effective ranges for these two systems, which are listed in Table~\ref{Tab:SL_ER}. The results from other three phenomenological models are also shown for comparison. The Rel. ChEFT results are calculated with the LECs listed in Tables~\ref{Tab:LECs_m3} and \ref{Tab:LECs_m4}. The corresponding predictions imply that the $^1S_0$ potentials of both $\Xi\Sigma$ ($I = 3/2$) and $\Xi\Xi$ ($I = 1$) are weakly attractive, and with the increase of strangeness, the attraction becomes even weaker. For the $^3S_1$ partial wave, the $\Xi\Sigma$ interaction is moderately attractive while the $\Xi\Xi$ interaction becomes repulsive.

\begin{table}[h]
  \centering
  \caption{$\Xi\Sigma$ ($I = 3/2$) and $\Xi\Xi$ singlet and triplet $S$-waves scattering lengths $a$ and effective ranges $r$ (in units of ${\rm fm}$) for various cutoff values $\Lambda_ F$ (in units of MeV). The last three columns show the results of the SU(6) quark cluster model (fss2)~\cite{Fujiwara2007ProgPartNuclPhys58.439} and the Nijmegen potentials (NSC97a, NSC97f)~\cite{Stoks1999PhysRevC59.3009}. }
  \label{Tab:SL_ER}
  \setlength{\tabcolsep}{15.9pt}
  \begin{tabular}{ccrrrrcrrr}
    \hline
    \hline
    \multicolumn{1}{c}{\multirow{2}{*}{}}&\multicolumn{5}{c}{Rel. ChEFT}&\multicolumn{1}{c}{\multirow{2}{*}{}}&\multicolumn{1}{c}{\multirow{2}{*}{fss2}}&\multicolumn{1}{c}{\multirow{2}{*}{NSC97a}}&\multicolumn{1}{c}{\multirow{2}{*}{NSC97f}}\\
    \cline{3-6}
    &&$550$ &$600$ &$650$ &$700$ && & &\\
    \hline
    $a_s^{\Xi\Sigma}$  &&$-1.02$ &$-1.02$ &$-1.02$ &$-1.02$ &&$-4.63$ &$4.13$ &$2.32$  \\
    $r_s^{\Xi\Sigma}$  &&$ 0.88$ &$ 0.91$ &$ 0.93$ &$ 0.94$ &&$ 2.39$ &$1.46$ &$1.17$  \\
    $a_t^{\Xi\Sigma}$  &&$-1.58$ &$-1.60$ &$-1.61$ &$-1.62$ &&$-3.48$ &$3.21$ &$1.71$  \\
    $r_t^{\Xi\Sigma}$  &&$ 2.10$ &$ 2.13$ &$ 2.15$ &$ 2.17$ &&$ 2.52$ &$1.28$ &$0.96$  \\
    \hline
    $a_s^{\Xi\Xi}$  &&$-0.46$ &$-0.46$ &$-0.46$ &$-0.46$ &&$-1.43$ &$17.28$ &$2.38$ \\
    $r_s^{\Xi\Xi}$  &&$ 7.24$ &$ 7.23$ &$ 7.20$ &$ 7.17$ &&$ 3.20$ &$ 1.85$ &$1.29$ \\
    $a_t^{\Xi\Xi}$  &&$ 0.16$ &$ 0.16$ &$ 0.17$ &$ 0.17$ &&$ 3.20$ &$ 0.40$ &$0.48$ \\
    $r_t^{\Xi\Xi}$  &&$11.07$ &$ 9.66$ &$ 8.74$ &$ 8.12$ &&$ 0.22$ &$ 3.45$ &$2.80$ \\
    \hline
    \hline
  \end{tabular}
\end{table}

\subsection{Evolution of the ``$27$'' irreducible representation in two-octet-baryon interactions}

Up to now we have systematically studied two-octet-baryon interactions in the relativistic ChEFT at leading order. Our results show that the $NN$ ($I = 1$) $^1S_0$ interaction~\cite{Ren2018ChinPhysC42.014103} is strongly attractive to generate a virtual bound state, the $\Sigma N$ ($I = 3/2$) $^1S_0$ interaction~\cite{KaiWen2018ChinPhysC42.014105} is moderately attractive, and the $\Sigma\Sigma$ ($I = 2$)~\cite{KaiWen2018PhysRevC98.065203}, $\Xi\Sigma$ ($I = 3/2$) and $\Xi\Xi$ ($I = 1$) $^1S_0$ interactions are weakly attractive. All of these five systems belong to the same SU(3) irreducible representation ``$27$''. Ideally the behaviors of these five states should be the same under strict SU(3) symmetry, but in practice  SU(3) symmetry is broken due to the mass difference of octet baryons and pseudoscalar mesons. Thus it offers an ideal place to understand SU(3) symmetry breaking via the evolution of the irreducible representation ``$27$''. We list the scattering lengths of these  five systems in Table~\ref{Tab:SC04}. The predictions from the three phenomenological models, i.e., fss2, NSC97a, and NSC97f, are also listed for comparison. Our results show that the attraction decreases fast as strangeness increases from $S = 0$ to $S = -2$, but then remains almost unchanged until $S = -4$. In particular, the $\Sigma\Sigma$ interaction is even less attractive than that of the $\Xi\Sigma$ system. It is worthwhile  emphasizing that the LECs in the $S = 0$ and $-1$ baryon-baryon potentials are determined by fitting to experimental data, while those from $S = -2$ to $-4$ are fitted to LQCD data. It is also interesting to note that the scattering lengths of the $\Sigma\Sigma$ ($I = 2$) channel are rather different in these models, and the result of the SU(6) quark cluster model will increase to $-9.27~{\rm fm}$ after taking into account the Coulomb force~\cite{Fujiwara2007ProgPartNuclPhys58.439}.

\begin{table}[h]
  \centering
  \caption{Singlet scattering lengths $a_s$ (in units of ${\rm fm}$) of the baryon–baryon systems from $S = 0$ to $S = -4$. The results are calculated  in the relativistic ChEFT at LO with a cutoff $\Lambda_F = 600$ MeV. The last three columns show results for the SU(6) quark cluster model (fss2)~\cite{Fujiwara2007ProgPartNuclPhys58.439} and the Nijmegen potentials (NSC97a, NSC97f)~\cite{Stoks1999PhysRevC59.3009}. Note that the Coulomb force is considered for the $\Sigma\Sigma$ ($I = 2$) channel in the NSC97a and NSC97f potentials, while not in the SU(6) quark cluster model and our present study.}
  \label{Tab:SC04}
  \setlength{\tabcolsep}{21pt}
  \begin{tabular}{lccrrrr}
  \hline
	\hline
  \multicolumn{1}{c}{\multirow{2}{*}{}}&\multicolumn{1}{c}{\multirow{2}{*}{Channel}}&\multicolumn{1}{c}{\multirow{2}{*}{Isospin}}&\multicolumn{4}{c}{$a_s$}\\
  \cline{4-7}
         &               &      &Rel. ChEFT &fss2     &NSC97a   &NSC97f    \\
	\hline
	$S=0$  &$NN$           &$1$   &$-21.30$   &$-23.76$ &$-15.84$ &$-14.49$  \\
	$S=-1$ &$\Sigma N$     &$3/2$ &$-4.04$    &$-2.51$  &$-6.06$  &$-6.16$   \\
	$S=-2$ &$\Sigma\Sigma$ &$2$   &$-0.80$    &$-85.30$ &$10.19$  &$6.98$    \\
	$S=-3$ &$\Xi\Sigma$    &$3/2$ &$-1.02$    &$-4.63$  &$ 4.13$  &$2.32$    \\
	$S=-4$ &$\Xi\Xi$       &$1$   &$-0.46$    &$-1.43$  &$17.55$  &$2.38$    \\
	\hline
	\hline
  \end{tabular}
\end{table}

\section{Summary and Outlook}

In this work, we have studied the strangeness $S = -3$ and $-4$ baryon–baryon interactions in the relativistic ChEFT at leading order. In order to be self-consistent, we first redetermined the hyperon-nucleon interactions by fitting the 12 LECs to the 36 $YN$ scattering data with cutoff $\Lambda_F = 550 - 700$ MeV. By assuming strict SU(3) flavor symmetry, the $S = -1$ baryon-baryon interactions were extended to the $S = -3$ and $-4$ sectors. Compared to the state-of-the-art LQCD simulations, it is found that there is an appreciable SU(3) flavor symmetry breaking from strangeness $S = -1$ to $S = -3$ and $-4$ sectors. In order to consider these effects, we redetermined two sets of LECs by fitting to the LQCD data for the $\Xi\Sigma$ and $\Xi\Xi$ channels respectively. The fitting results demonstrated that the $S$-waves phase shifts of LQCD can be described rather well. In addition, without any additional free LECs, the predicted phase shifts for the $^3D_1$ channel and the mixing angle $\varepsilon_1$ are also found to be in qualitative agreement with the LQCD data in the $S = -3$ sector. Quantitatively, the relativistic ChEFT predicted a stronger coupling than the LQCD for the $^3S_1-{}^3D_1$ channel. With the obtained LECs, the $S$-wave scattering lengths and effective ranges were also calculated for these two systems at the physical point. Finally, using the $S = 0$ and $-2$ results obtained in our previous works, we studied the evolution of the singlet $S$-wave scattering lengths with the increase of strangeness, which belong to the same irreducible representation ``$27$''. It was shown that the attraction decreases fast as strangeness increases from $S = 0$ to $S = -2$, but then remains almost unchanged from $S = -2$ to $S = -4$, which indicates that the existence of bound states in the $\Xi\Sigma$ and $\Xi\Xi$ systems is rather unlikely.

Although experimental studies of baryon-baryon interactions of strangeness $S = -3$ and $-4$ are rather challenging, it has been recently demonstrated that it is possible to extract information on such interactions from correlation measurements in heavy-ion collisions at RHIC or CERN~\cite{Adamczyk2015PRL114.022301}. In addition, baryon-baryon systems with $S = -2,-3,-4$ can also be produced in photon induced reactions on the deuteron at JLab~\cite{Miller2013CJP51.466} or $K^-$ induced reactions at J-PARC~\cite{Ahn2011JPARC}. In the not so far future, these studies might substantially advance our understanding of multi-strangeness systems.

\section*{Acknowledgements}
This work is partly supported by the National Natural Science Foundation of China under Grant Nos.11735003, 11975041, 11775148, and 11961141004.

\bibliography{BB.bib}
\end{document}